\documentclass[graybox]{svmult}

\usepackage{type1cm}        
\usepackage{makeidx}         
\usepackage{graphicx}        
\usepackage{multicol}        
\usepackage[bottom]{footmisc}

\usepackage[varvw]{newtxmath}       
\usepackage[T1]{fontenc}
\usepackage[utf8]{inputenc}

\makeindex             

\begin{document}

\title*{Unsupervised classification reveals new evolutionary pathways}
\titlerunning{ML applied to galaxy classification}
\author{M. Siudek\inst{1,2}, K. Lisiecki\inst{3,4}, M. Mezcua\inst{1,5}, K. Małek\inst{3}, A. Pollo\inst{3,6}, J. Krywult\inst{7}, A.~Karska\inst{4,8}, Junais\inst{3}}
\authorrunning{M. Siudek et. al}
\institute{${1.}$ Institute of Space Sciences (ICE, CSIC), Campus UAB, Carrerde Can Magrans, s/n, 08193 Barcelona, Spain ${2.}$ Institut de Física d’Altes Energies (IFAE), The Barcelona Institute of Science and Technology, 08193 Bellaterra (Barcelona), Spain ${3.}$ National Centre for Nuclear Research, Pasteura 7, 093, Warsaw, Poland 
${4.}$ Institute of Astronomy, Faculty of Physics, Astronomy and Informatics, Nicolaus Copernicus University, Grudzi{\k a}dzka 5, 87-100 Toru{\'n}, Poland ${5.}$ Institut d'Estudis Espacials de Catalunya (IEEC), Carrer Gran Capit\`a, 08034 Barcelona, Spain ${6.}$ Astronomical Observatory of the Jagiellonian University, ul. Orla 171, 30-244 Kraków, Poland 
${7.}$ Institute of Physics, Jan Kochanowski University, ul. Uniwersytecka 7, 25-406 Kielce, Poland ${8.}$ Max-Planck-Institut für Radioastronomie, Auf dem Hügel 69, 53121, Bonn, Germany;
\email{msiudek@ifae.es}}
%
%
\maketitle
\abstract{While we already seem to have a general scenario of the evolution of different types of galaxies, a complete and satisfactory understanding of the processes that led to the formation of all the variety of today’s galaxy types is still beyond our reach. To solve this problem, we need both large datasets reaching high redshifts and novel methodologies for dealing with them. The VIPERS survey statistical power, which observed $\sim90,000$ galaxies at $z>0.5$, and the application of an unsupervised clustering algorithm allowed us to distinguish 12 galaxy classes. 
Studies of their environmental dependence indicate that this classification may actually reflect different galaxy evolutionary paths. For instance, a class of the most passive red galaxies gathers galaxies $\sim20\%$ smaller than other red galaxies of a similar stellar mass, revealing the first sample of red nuggets at intermediate redshift. On the other end, a class of blue dwarf galaxies is composed mainly of AGN, challenging commonly used mid-infrared AGN selections. 
}

\section{Introduction}\label{sec:introduction}

Nowadays it is essential to apply efficient classification procedures to process large astronomical datasets to understand the universe and its evolution. 
Modern deep surveys, such as the Dark Energy Spectroscopic Instrument (DESI; \cite{Abareshi2022}) will gather spectra of millions of objects among an even more unprecedented set of over a billion galaxies from imaging surveys (DESI Legacy Imaging Surveys; \cite{Dey2019}). 
Standard approaches to galaxy classification have relied on a manual or citizen scientists' classification,  
however, such approaches do not scale well with the unprecedented data-growth rate.  
The solution is the application of ground-breaking machine learning (ML) and deep learning algorithms.  
Unsupervised (without any training process) ML techniques, which can extract new knowledge and identify so far unknown populations of objects, are of particular interest  and have already shown their potential in discovering new galaxy pathways 
(see~\cite{Baron2019} for review). 
In this work, we discuss the potential of the unsupervised ML Fisher Expectation-Maximization (FEM) algorithm~\cite{Bouveyron2011} in revealing peculiar classes of the galaxy population. 

\section{Data}\label{sec:data}
The data used in this proceeding comes from the VIMOS Public Extragalactic Redshift Survey (VIPERS;~\cite{Scodeggio2018}). 
Observations were performed with the VIsible MultiObject Spectrograph 
mounted on the Very Large Telescope at ESO. 
The survey covers $\rm{\sim23.5\mbox{ } deg^2}$ split over two W1 and W4 fields of the Canada-France-Hawaii Telescope Legacy Survey Wide. 
The final data release consists of 86,775 sources limited to $\rm{i_{AB} \leq 22.5}$ mag targeting galaxies at redshift range $0.4\lesssim z \lesssim 1.2$. Additionally, the VIPERS survey observed a sample of broad-line AGN ($\sim1\%$ of the total VIPERS sample) reaching redshift up to  $z\sim4$. 

\section{Unsupervised galaxy classification}\label{sec:results}
\begin{figure}
	\centerline{\includegraphics[width=0.5\textwidth]{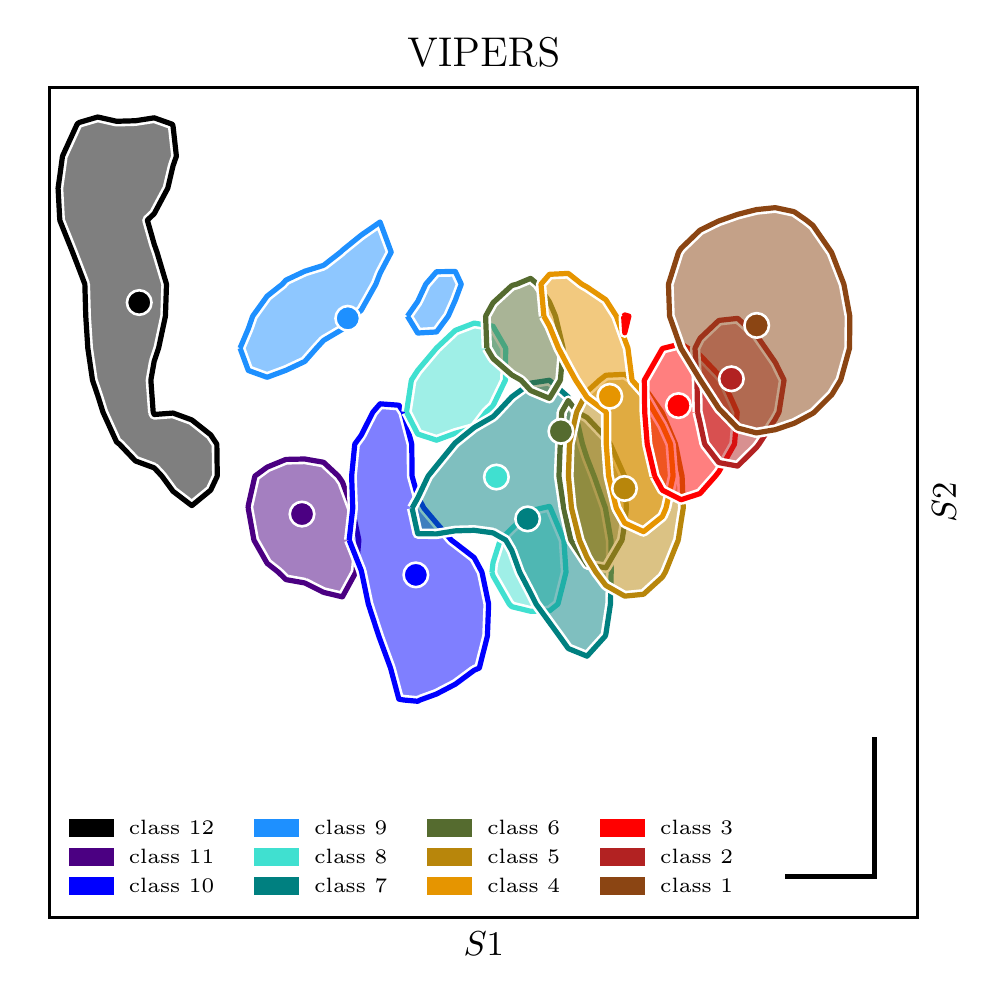}}
	\caption{Projections of VIPERS samples from~\cite{Siudek2018a} onto the two dimensions that best separate their clusters. The axes of each plot are determined by FEM and are not corresponding to any peculiar galaxy parameter. }
	\label{fig:fem}
\end{figure}

VIPERS galaxies with the highest redshift confidence level (> 90\%) were separated using the FEM algorithm into 12 clusters corresponding to 3 red, 3 green, and 5 blue galaxy classes and one of the broad-line AGN by~\cite{Siudek2018a}. 
The clustering algorithm works in a 13-dimensional feature space composed of spectroscopic redshift and 12 rest-frame ultraviolet-through-near-infrared magnitudes normalised to the i-band magnitude. 
Figure~\ref{fig:fem} shows the projection of the VIPERS sample onto the two principal dimensions of the respective 12-dimensional latent subspace. 
The FEM classification was also successfully applied on solely photometric datasets (i.e. using photometric redshifts) achieving high accuracy ($\sim90\%$) of reproducing spectroscopic classification as well as high efficiency ($>90\%$) in the separation of stars and broad-line AGN~\cite{Siudek2018b}. 
A multi-epoch approach based on the application of the FEM algorithm to VIPERS ($z\sim0.7$) and SDSS ($z\sim0.0$) data allowed us to study the cosmic evolution of galaxy classes, which we find to mostly reflect the gradual internally-driven growth of bulges and slow quenching~\cite{Turner2021}. 

\begin{figure}
	\centerline{\includegraphics[width=0.99\textwidth]{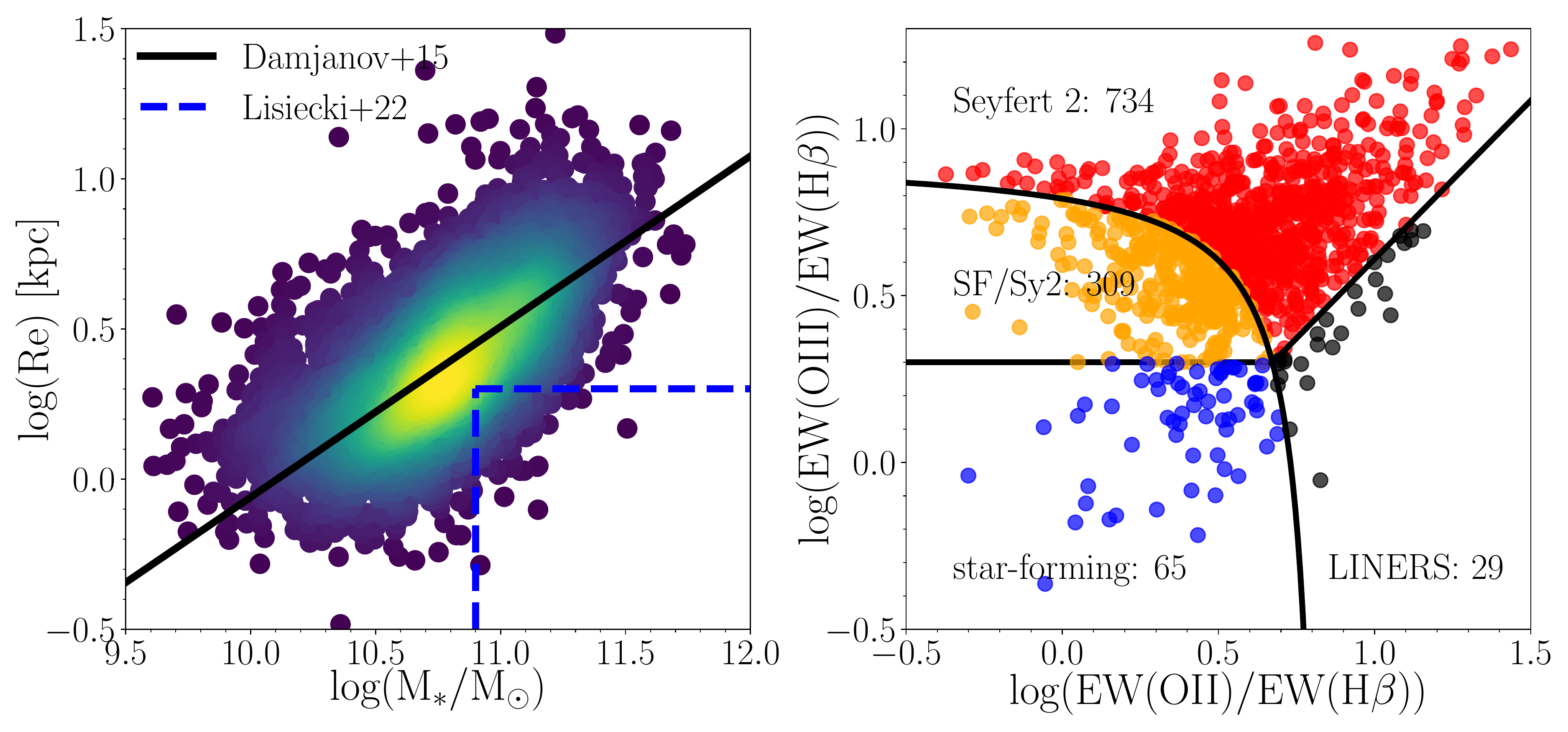}}
	\caption{Left panel: The compactness diagram for the most passive VIPERS class which gathers the first intermediate-redshift sample of red nuggets~\cite{Lisiecki2022,Siudek2022b} among UCMG (with a cut proposed by~\cite{Damjanov2015}). Right panel: The emission line diagnostic diagram~\cite{Lamareille2010} for the bluest VIPERS class which gathers a population of dwarf galaxies hosting AGN~\cite{Siudek2022c}.  }
	\label{fig:figure1}
\end{figure}

\paragraph{Red compact massive galaxies}\label{sec:rednuggets}
The FEM algorithm distinguished 3 red classes, which catch a part of the rare population of red ultra-compact massive galaxies (UCMG) that after formation at high redshift ($z>2$) have avoided merger processes (red nuggets). 
In fact, less than 100 UCMG were so far observed at $z<0.5$~\cite{Scognamiglio2020} leaving the intermediate redshift ($0.5<z<2.0$) unexplored. VIPERS recently filled this niche by providing an unprecedented sample of 77 red nugget candidates~\cite{Lisiecki2022,Siudek2022b}, of which 72 (94\%) are found within FEM red classes. In particular, the majority of them (65\%) are found in class~1, containing the most massive and the smallest red galaxies~\cite{Siudek2022a}. 
The red nugget catalogue is created with the most restricted selection cuts (in stellar mass and effective radius, see the left panel in Fig.~\ref{fig:figure1}), however, class~1 gathers also larger/less massive galaxies (but still smaller and more massive than standard passive galaxies~\cite{Siudek2017}) further inviting us to explore UCMG in states of their transition.

\paragraph{AGN dwarf galaxies}\label{sec:dwarfAGN}
Among blue galaxies, the FEM algorithm identified galaxies with the bluest $NUVr$ and $rK$ colours (class~11), which contain AGN~\cite{Siudek2018a,Siudek2022c}. 
These are in majority (90\%) dwarf galaxies ($\rm{log(M_{*}/M_{\odot})<9.5}$). 
Their distribution on the emission line diagnostic diagram (see the right panel in Fig.~\ref{fig:figure1}) confirms that this class is built mainly of either AGN (67\%) or composite systems (27\%\footnote{It is expected that $\sim20\%$ of composite objects are AGN. Thus, it is possible that our classification successfully distinguishes AGN and non-AGN galaxies located in the composite region.}) and a small fraction of star-forming galaxies (6\%). 
At the same time, 27\% of all spectroscopically-identified VIPERS AGN are gathered in class~11. 
Such a performance of automatic AGN dwarf selection (accuracy of at least $67\%$ and completeness of 27\%) 
challenges standardly used mid-infrared AGN selection (with an accuracy of $\sim60\%$ and completeness of $\sim15\%$, see Tab. 6 in~\cite{Hviding2022}). 
Moreover, AGN in dwarf galaxies are nowadays of special interest as most of them are believed to host intermediate-mass black holes, the seeds of supermassive black holes~\cite{Siudek2022c,Mezcua2022}. 


\section{Summary}\label{sec:summary}
The unsupervised colour-based VIPERS classification~\cite{Siudek2018a} 
shows great potential for studying physical mechanisms regulating galaxy evolution and automatically revealing classes of peculiar galaxies. In particular, we show that: 
\begin{itemize}
    \item large photometric samples can be used to distinguish different galaxy classes with an accuracy similar to spectroscopic classification~\cite{Siudek2018b},
    \item different evolutionary paths of galaxies are correlated with their environments~\cite{Siudek2022a},    
    \item galaxy classes at two different epochs are correlated with each other~\cite{Turner2021},
    \item we are able to automatically identify classes of peculiar objects: i) the first intermediate-redshift sample of UCMG~\cite{Lisiecki2022,Siudek2022b, Siudek2022a}, ii) AGN dwarf galaxies with a performance outshining mid-infrared AGN selections~\cite{Siudek2022c}.
\end{itemize}

\begin{acknowledgement}
This work has been supported by the Polish National Agency for Academic Exchange (Bekker grants BPN/BEK/2021/1/00298/DEC/1 and BPN/BEK/2021/1/00319/DEC/1), the European Union's Horizon 2020 Research and Innovation Programme under the Maria Sklodowska-Curie grant agreement (No. 754510), and the Spanish Ministry of Science and Innovation through the Juan de la Cierva-formacion programme (FJC2018-038792-I), 
the Polish NCN (UMO-2018/30/E/ST9/00082)  and the Ramon y Cajal fellowship (RYC2019-027670-I). This work was also partially supported by the program Unidad de Excelencia Mar\'ia de Maeztu CEX2020-001058-M.

\end{acknowledgement}

%
\newcommand*\aap{A\&A}
\let\astap=\aap
\newcommand*\aapr{A\&A~Rev.}
\newcommand*\aaps{A\&AS}
\newcommand*\actaa{Acta Astron.}
\newcommand*\aj{AJ}
\newcommand*\ao{Appl.~Opt.}
\let\applopt\ao
\newcommand*\apj{ApJ}
\newcommand*\apjl{ApJ}
\let\apjlett\apjl
\newcommand*\apjs{ApJS}
\let\apjsupp\apjs
\newcommand*\aplett{Astrophys.~Lett.}
\newcommand*\apspr{Astrophys.~Space~Phys.~Res.}
\newcommand*\apss{Ap\&SS}
\newcommand*\araa{ARA\&A}
\newcommand*\azh{AZh}
\newcommand*\baas{BAAS}
\newcommand*\bac{Bull. astr. Inst. Czechosl.}
\newcommand*\bain{Bull.~Astron.~Inst.~Netherlands}
\newcommand*\caa{Chinese Astron. Astrophys.}
\newcommand*\cjaa{Chinese J. Astron. Astrophys.}
\newcommand*\fcp{Fund.~Cosmic~Phys.}
\newcommand*\gca{Geochim.~Cosmochim.~Acta}
\newcommand*\grl{Geophys.~Res.~Lett.}
\newcommand*\iaucirc{IAU~Circ.}
\newcommand*\icarus{Icarus}
\newcommand*\jcap{J. Cosmology Astropart. Phys.}
\newcommand*\jcp{J.~Chem.~Phys.}
\newcommand*\jgr{J.~Geophys.~Res.}
\newcommand*\jqsrt{J.~Quant.~Spectr.~Rad.~Transf.}
\newcommand*\jrasc{JRASC}
\newcommand*\memras{MmRAS}
\newcommand*\memsai{Mem.~Soc.~Astron.~Italiana}
\newcommand*\mnras{MNRAS}
\newcommand*\na{New A}
\newcommand*\nar{New A Rev.}
\newcommand*\nat{Nature}
\newcommand*\nphysa{Nucl.~Phys.~A}
\newcommand*\pasa{PASA}
\newcommand*\pasj{PASJ}
\newcommand*\pasp{PASP}
\newcommand*\physrep{Phys.~Rep.}
\newcommand*\physscr{Phys.~Scr}
\newcommand*\planss{Planet.~Space~Sci.}
\newcommand*\pra{Phys.~Rev.~A}
\newcommand*\prb{Phys.~Rev.~B}
\newcommand*\prc{Phys.~Rev.~C}
\newcommand*\prd{Phys.~Rev.~D}
\newcommand*\pre{Phys.~Rev.~E}
\newcommand*\prl{Phys.~Rev.~Lett.}
\newcommand*\procspie{Proc.~SPIE}
\newcommand*\qjras{QJRAS}
\newcommand*\rmxaa{Rev. Mexicana Astron. Astrofis.}
\newcommand*\skytel{S\&T}
\newcommand*\solphys{Sol.~Phys.}
\newcommand*\sovast{Soviet~Ast.}
\newcommand*\ssr{Space~Sci.~Rev.}
\newcommand*\zap{ZAp}
%
%

\end{document}